\documentclass[aps,prd,reprint,showpacs,showkeys,amsmath,amssymb,amsfonts]{revtex4-1}
\pdfoutput=1
\usepackage{blindtext}
\usepackage{graphicx}
\usepackage{fullpage}
\usepackage{dcolumn}
\usepackage{tabularx}
\usepackage{natbib}
\usepackage{mathtools}

\usepackage[x11names]{xcolor}
\usepackage{hyperref}
\hypersetup{
	allbordercolors={blue},
	colorlinks=false, 
	citecolor={blue},	 
	citebordercolor={blue},
	linktoc=all,	 
	linkcolor={blue},	 
	urlbordercolor={Gold1}
}

\begin{document}
\title{Viscoelastic Theory Representation Of Gravitational Strain Fields In The
${}^+\Lambda$-CDM Vacuum}
\author{Ronald S. Gamble, Jr}
\email{ron\_gamble1@yahoo.com}
\author{K.M.Flurchick}
\affiliation{Department of Computational Sciences \& Engineering\\North Carolina
Agricultural \& Technical State University\\Greensboro, NC, 27401}

\begin{abstract}
In recent years, investigations of gravitational interactions has led us to
discover new facets of the fundamental force. With these discoveries the general
theory of relativity is under greater scrutiny now than it was 100 years ago.
Development of a more advanced theory is needed for experimental tests and
extensive predictions of gravitational interactions in local and cosmological
settings. In this work, we present axioms of a novel theory that describes
spacetime as a relativistic viscoelastic continuum; retaining the properties of the natural
${}^{+}\Lambda$-CDM vacuum. With these axioms, we provide a foundation
for a tensor field theory of Gravitational Strain; introduced as an advanced
formulation of an elastic interpretation of gravitational interactions. 
This viscoelastic definition of gravity is a natural advancement of general
relativity to an observable measure of a proper relativistic
gravitational field.
\end{abstract}


\maketitle

\section*{Introduction}
Since the genesis of experimental tests of general relativity, the field of
astrophysics has slowly veered away from being solely an ``observational''
science.
Experimental astrophysics is emerging as a propagator of novel and innovative
constructs that helps to transform concepts in theoretical physics into tangible
tests of physical processes. This builds a long sought after bridge between
theory and experiment that has been considered elusive in regards to testing any
relativistic theory of gravity.
During the last few years the largest experimental astrophysics project to date,
the LIGO/LSC/VIRGO collaboration, has been a champion of this effort in bridging
theory with experiment. Experimentation spearheads and propels theoretical
concepts and predictions into a broader consensus. The recent activities by this
major collaboration has given motivation to redefine what we interpret
the \textit{fabric} of spacetime as. Einstein and his contemporaries insisted
that a much more advanced theory is needed to describe the \textit{strangeness}
that is observed in the universe (e.g. black holes, wormholes, and ``dark''
objects). With this being said, the
development of a more advanced theory based on observations of nature is needed
for experimental tests of gravitational interactions in a local and cosmological
setting.

In this work we introduce an innovative approach to bridge theory and
experimentation by presenting a novel interpretation of spacetime as a relativistic viscoelastic fluid. Considering the results of the first LIGO
observations \cite{ligo,ligo2}, it is logical to hypothesize an elastic nature
to the interactions of spacetime geometry and the measure associated with the
gravitational field; strain. Considering this interpretation of spacetime
geometry in terms of \textit{gravitational strain} provides for a more
physical interpretation of the properties of the gravitational field
interactions and subsequent propagation through spacetime. This can be seen by
the observations of gravitational strain in the gravitational wave astronomy
community. Thus, a diversion from Riemannian geometry and general relativity is
not needed. The geometric constructs in Riemannian geometry supports this
extension of general relativity. A Gravitational Strain Field Theory finds
itself at the junction of the metric-theory general relativity and the results
from contemporary methodologies (LISA, and LIGO\cite{lisa,ligo2,ligo}), and
observations (Hulse-Taylor Binary Pulsar \cite{pulsar}).

Since Newton's \textit{Philosophi\ae} \textit{Naturalis Principia Mathematica}
in 1687 we perceive the fundamental force of Gravity as having \textit{static}
properties; lacking a time-dependent nature to its field equations and resulting
interactions.
Even with the rise of Einstein's remarkable metric theory\cite{efe} of gravity
in 1916 and its numerous extensions and interpretations, a dynamic theory of
gravitation has eluded physicists to date. Given that the
general theory of relativity is a metric theory of gravity, and that the metric
tensor is not a direct observable \cite{nonlinear,pulsar2}, it is an obvious
approach to consider spacetime strain as an observable that would have a
strongly correlated relationship with the metric of spacetime. A closer
examination of the formulations in continuum mechanics shows that Einstein's
relativity can be considered as one interpretation and application of a
generalized theory concerning dynamics of a \textit{continuum}.
An elastic definition of gravitational interactions would be a natural extension
of general relativity to an observable measure of a proper gravitational field.

By considering a covariant formulation with the axioms defined in section
\ref{axiom} we demonstrate that a generalized equation can be described to
permit this proposed elastic nature of spacetime. This formulation which is
imposed upon an ambient \textit{dynamic} background which includes a new
approach to using a combination of superimposed metrics to construct a
background measure.
The geometrical representation of a gravitational field, generated by a mass
distribution, is thus described as a set of coupled elastic deformations of the
distribution. This is stated with respect to variations in the volume, surface
shape, and rotation. The respective tensor fields are the \textit{Dilation}
($\mathcal{D}$), \textit{Shear-Tidal} ($\mathcal{S}$) and \textit{Vorticity
Induction} ($\mathcal{V}$) of the source distribution. The concentration of this
work is on the formulation of the two tensor fields $\mathcal{D}$ and $\mathcal{S}$, excluding
the rotational contributions.
Rotational attributes of the gravitational strain field will be addressed later,
but is mentioned here for the completion of the strain fields. Terms
corresponding to deformations like that of a covariant strain tensor \(\left(\varepsilon _{\mu \nu }\right)\) and cosmological Bulk
Incompressibility ($\beta$) will be used in the formulation of the two
respective strain fields in section \ref{fields}.

\section{\label{forms}Strong and Weak Background Metric Forms}

Traditionally, when one is solving the Einstein field equations a
\textit{single} metric solution is sought after that supports a description of
the spacetime structure for a singular static source of mass-energy.
This metric solution generally gives a linear measure of distance and causality
within the local spacetime under the influence of said mass-energy source. If we
consider that this (matter) source distribution is contained within an ambient
\textit{environment}, with respective energetic properties (nonzero energy
density), then there must also exist a metric solution to account for this
ambient geometry and properties.
Thus, interpreting the local spacetime environment surrounding a matter source
is described by a composition of metric solutions representing strong and weak
matter sources. This is done by imposing a non-static background that includes
using a superposition of cosmological and matter metric solutions to construct a
composite background measure. 

Strong and weak forms are constructed by an imposed
combination of \textit{local} and
\textit{cosmological}
(Fr$\ddot{e}$idman-L$\hat{a}$timere-Robertson-Walker, FLRW metric)
\cite{mtw,spacetime} spacetime metrics with the $\Lambda $-Vacuum metric.
The \textit{strong} and \textit{weak} form background metrics are expressed as
external direct sums ($\oplus$) of the characteristic parts pertaining to the
spacetime background. As can be seen, the strong form metric contains the
combination of the $\Lambda$-vacuum (${}^+\Lambda\eta_{\mu\nu}$) with the local
geometric measure of the spacetime associated with a matter distribution
($\chi_{\mu\nu}$).
\begin{equation}\label{eq:strong}
\Lambda _{\mu \nu }{}^{(\text{background})}= {}^+\Lambda \eta _{\mu \nu
}{}^{(\Lambda -\text{Vacuum})} \oplus  \chi _{\mu \nu }{}^{(\text{source})}
\end{equation}
with the \textit{weak form} background metric tensor as
\begin{eqnarray}\label{eq:weak}
 \lambda _{\mu \nu }{}^{(\text{background})}&=& {}^+\Lambda \eta _{\mu \nu
 }{}^{(\Lambda -\text{Vacuum})}\\ 
 &\oplus& e^{\alpha (\tau )}\cdot \chi _{\mu \nu
 }{}^{(\text{source})}\oplus f_{\mu \nu }{}^{(\text{FLRW})}\nonumber
\end{eqnarray}
At spatial infinity from the source, the weak form includes the relative
expansion of spacetime pertaining to cosmological length scales, and subsequent
redshift. 

Determining which background form is appropriate for use within the local
spacetime structure is done by considering the proximity to
the matter distribution. When considering the volume or extension of a closed
surface surrounding the distribution, we can set a bounds on the extension of
the local field. Thus, determining which background form is appropriate for the
description of the spacetime structure. Equation (3) below represents a
generalization of these requirements on the background form choice.
\begin{eqnarray}
 g_{\mu \nu }\equiv
\begin{cases}
 \Lambda _{\mu \nu }{}^{(\text{strong})} & \left(\text{local to source}\right)
 \\
 \lambda _{\mu \nu }{}^{(\text{weak})} & \left(\text{far from source}\right)
 \\
\end{cases}
\end{eqnarray}
The source metric will take on a metric solution that is respective of
describing the geometry of a local distribution, in this representation of a
background metric tensor. The inclusion of the $\Lambda-$vacuum in both strong
and weak background measures is a statement of the zero state gravitational
vacuum for all space.
Excluding the contribution of curvature by sources of matter, this is the
fundamental metric of the vacuum with nonzero energy density under this
representation.
As for the source metric, $\chi_{\mu\nu}$, this metric is left undefined. As it
can take on representation of any metric solution corresponding to the
Einstein Field Equations for confined masses and compact objects (e.g. Kerr, Schwarzschild, Kerr-Newman,
etc.).

\section{\label{axiom}The Viscoelasticity of Spacetime}
\subsection{Axioms of Viscoelastic Theory}

The following axioms are defined for a background spacetime represented as a
\textit{relativistic Viscoelastic fluid}:
\begin{enumerate}
\item Physical spacetime is described as a smooth continuous Pseudo-Riemannian
manifold (Lorentzian manifold), with appropriate rotational and
translational symmetries.
The spacetime vacuum can be assumed to be {``}fluid-like{''} because of its attributes as a smooth
continuous manifold, satisfying the symmetries of relativity.
\item The Lorentzian manifold is described by a metric tensor solution
\(\left(\Lambda_{\mu \nu }\right)\). Due to Lorentz transformations of the
spacetime coordinates, the resulting length contraction and time dilation hold along with a deformable
{``}local{''} spacetime volume.
\item A defined flow of energy-momentum through the vacuum is expressed in terms
of a defined set of current densities $\{\mathcal{J}_{\mu}\}_{\alpha}$.
\item An intrinsic zero state compressibility/expansion arises in the form of a
nonzero vacuum pressure, the Cosmological Constant, ${}^+\Lambda$. Taken
for positive values of $\Lambda$ (not to be confused with the strong-form
background metric tensor, $\Lambda_{\mu\nu}$, as mentioned above.).
\end{enumerate}

With these assumptions of the vacuum fluidity, the background spacetime can
naturally give rise to dynamic properties of elasticity in
n-dimensions. The following formulation of the fundamental properties of
continuum elasticity is given as derived in the text \cite{elastic1}.
Applications of this description of elasticity serves to provide a fundamental basis from which
a viscoelastic theory of spacetime can be formulated. From this, an analogy of
viscoelasticity can be applied towards interpreting gravitational field
interactions.

\subsection{Formulation of Spacetime Viscoelasticity}
Firstly, a notion of the comparison of reference and deformed configurations
is made. Providing a statement of the material deformation and unit volume
expansion/compression. The \textit{Lagrangian}
(material) Deformation Gradient Tensor $\Delta _{\hat{\mu} \nu }$, as a
\textit{two-point tensor}, characterizes the local deformation at a continuum point, with local coordinates \(\left(x_{\mu }\right)\) described by
\begin{equation}
\Delta _{\hat{\mu }\nu } = \Delta \left(e_{\hat{\mu }}\otimes I_{\nu }\right)
\end{equation}
The deformation at neighboring points is given by transferring a continuum line
element (ds) emanating from that point in the reference configuration (labeled
by indices without a hat) to the current deformed ($\hat{\mu }$)
configuration. With continuity condition, the mapping function $u\left(x_{\mu
}\right)$, is a function of the local coordinates.
\begin{equation}
\partial _{\beta }u\left(x_{\alpha }\right)dx_{\alpha } = \Delta _{\alpha \beta }dx_{\alpha }
\end{equation}
where, 
\begin{equation}
\partial _{\alpha }u_{\beta }\left(x_{\alpha }\right) \equiv  \frac{du_{\hat{\beta }}}{dx_{\alpha }}= \Delta _{\alpha \hat{\beta }}
\end{equation}
describes the {``}motion{''} of the continuum. The metric in the reference
configuration has components, \(g_{\mu \nu }\left(\partial _{(n)}\right) =
g\left(\partial _{\mu },\partial _{\nu }\right)\); and relative to its new
configuration in the deformed coordinate system \(\left(u_{\hat{\nu
}}\left(x_{\hat{\mu }}\right)\right)\) the metric will determine a different
tensor of coefficients: \(g_{\hat{\mu }\hat{\nu }}\left(\partial
_{\left(\hat{n}\right)}\right) = g\left(\partial _{\hat{\mu }}, \partial
_{\hat{\nu }}\right).\) The new coordinates are related to the metric in the
reference configuration by the transformation.
\begin{equation}
g^{\hat{\mu }\hat{\nu }}[u]=\Delta ^{\hat{\mu }\alpha }g_{\alpha \beta }[x]\Delta ^{\beta \hat{\nu }}
\end{equation}
Where the line elements for each of the configurations (reference and deformed)
are:
\begin{equation}
 d\hat{s}^2=
g^{\hat{\mu }\hat{\nu }}[u]dU_{\hat{\mu }}dW_{\hat{\nu }}
\end{equation}
\begin{equation}
ds^2= g_{\alpha \beta }[x]dU^{\alpha }dW^{\beta }
\end{equation}
For the deformation gradient tensor, a notion of spacetime (continuum)
compressibility is made via the determinant and its relationship to the local
volume:
\begin{equation}
\det \left[\Delta _{\hat{\mu }\nu }\right]
\begin{cases}
 >1,  & \text{expansion} \\
 =1,  & \text{constant} \\
 <1,  & \text{compression} \\
\end{cases}
\end{equation}
Fundamentally, the deformation gradient tensor serves as a notion of
a background continuum transformation.

\subsection{Cauchy Deformation and Metric Equality}
In order to further define deformations in a background spacetime, a statement
of the \textit{metric compatibility} (describing the definition of deformed
paths and metric equality) must be made for the succeeding Cauchy deformation
tensor in 4-dimensions. On a Lorentzian manifold the Cauchy deformation tensor
$E_{\hat{\mu }\hat{\nu }}$ has the form stating,
\begin{eqnarray}
 E_{\hat{\mu }\hat{\nu }}&=&\Delta ^{\hat{\mu }\alpha }\Delta _{\alpha \hat{\nu
 }} \\
 &=& \frac{\partial u_{\alpha }}{\partial x_{\hat{\mu }}}\frac{\partial
 u_{\alpha }}{\partial x_{\hat{\nu }}}\nonumber
\end{eqnarray}
where the deformation tensor gives the square of the { }{``}local{''} change in
coordinate distances due to deformation:
\begin{equation}
du^2 = E_{\hat{\mu }\hat{\nu }}dx^{\hat{\mu }}dx^{\hat{\nu }}
\end{equation}
Such that changes in parametrized length are characterized by this deformation
tensor. Prior to deformation the path is found by the integral below defining
the parameterized path with $\lambda$ as an affine parameter. Before
deformation,
\begin{equation}
s[\lambda] = \int_0^\lambda \sqrt{\frac{dx_{\mu }}{d\lambda }\cdot \eta _{\mu
\nu }\cdot \frac{dx_{\nu }}{d\lambda }} \, d\lambda
\end{equation}
and after deformation
\begin{equation}
\hat{s}[\lambda] = \int_0^\lambda \sqrt{\frac{dx_{\mu }}{d\lambda }\cdot
E_{\hat{\mu }\hat{\nu }}\cdot \frac{dx_{\nu }}{d\lambda }} \, d\lambda
\end{equation}
giving a relative coordinate displacement of
$\Delta\tilde{\text{S}}=\hat{S}-S$.
With these definitions of the Cauchy deformation tensor an equality can be made
such that the deformation tensor behaves like that of a \textit{deformed} metric
tensor. In which it can be defined for the coordinate displacement field on a
smooth manifold with the required relationship, $E_{\hat{\mu }\hat{\nu }}\equiv
\tilde{g}_{\mu \nu }$, where $\tilde{g}_{\mu \nu }$ is the metric in the deformed configuration
with respect to the reference configuration coordinates.
From this relationship, an expression for defining an affine connection in the deformation
configuration can be given for such a metric equality,
\begin{equation}
\tilde{\Gamma }^{\hat{\alpha}}{}_{\hat{\gamma} \hat{\sigma} } =
\frac{1}{2}E^{\hat{\alpha }\hat{\lambda }}\left(\partial _{\hat{\sigma
}}E_{\hat{\gamma }\hat{\lambda }} + \partial _{\hat{\gamma }}E_{\hat{\lambda
}\hat{\sigma }} - \partial _{\hat{\lambda }}E_{\hat{\gamma }\hat{\sigma
}}\right)
\end{equation}
Naturally, we can then use the constructs of Riemannian geometry to
construct the resulting \textit{deformed} Riemann curvature tensor:
\begin{equation}
\tilde{R}_{\hat{\mu} \hat{\nu} \hat{\alpha} \hat{\beta}}= \partial _{\beta
}\tilde{\Gamma }^{\hat{\mu} }{}_{\hat{\nu} \hat{\alpha} } +
\tilde{\Gamma }^{\hat{\mu} }{}_{\hat{\gamma} \hat{\beta}
}\tilde{\Gamma}^{\hat{\gamma} }{}_{\hat{\mu} \hat{\nu} } - \partial
_{\alpha }\tilde{\Gamma }^{\hat{\mu} }{}_{\hat{\nu} \hat{\beta} } -
\tilde{\Gamma }^{\hat{\mu} }{}_{\hat{\gamma} \hat{\alpha}
}\tilde{\Gamma }^{\hat{\gamma} }{}_{\hat{\nu} \hat{\beta} }
\end{equation}
and Ricci curvature tensor
\begin{equation}
 \tilde{R}_{\hat{\mu}\hat{\nu} }=\partial _{\hat{\alpha} }\Gamma^{\hat{\alpha}
 }{}_{\hat{\mu} \hat{\nu} }-\partial_{\hat{\nu} }\Gamma^{\hat{\alpha}
 }{}_{\hat{\mu} \hat{\alpha} }+\Gamma^{\hat{\beta} }{}_{\hat{\mu} \hat{\beta}
 }\Gamma ^{\hat{\beta} }{}_{\hat{\alpha} \hat{\beta} }-\Gamma ^{\hat{\beta}
 }{}_{\hat{\mu} \hat{\alpha} }\Gamma ^{\hat{\alpha} }{}_{\hat{\nu} \hat{\beta} }
\end{equation}
This deformed Ricci tensor represents the amount of spacetime curvature
generated by the stress attributed to the deformation of the sourced local mass
distribution. 

When the background is represented by a zero-vacuum metric, for $\Lambda _{\mu
\nu }{}^{(\text{background})}\equiv\Lambda_{\mu\nu}{}^{(\text{zero})}$ or $\lambda
_{\mu \nu }{}^{(\text{background})}\equiv\lambda_{\mu\nu}{}^{(\text{zero})}$,
we can see that even in the absence of a mass (matter) distribution, the scalar
curvature is intuitively nonzero for this spacetime in the presence of a nonzero
vacuum energy density.
Now, when considering this zero-vacuum as somewhat of a zero state
configuration, the scalar curvature is inherently nonzero with a value of
$R_{vac}\equiv R_{zero}$, when $R_{zero} =
\Lambda^{\alpha\sigma}{}_{(\text{zero})}R_{\alpha\sigma}\equiv
\frac{2n}{n-2}({}^{+}\Lambda)$ for dimension (n = 4). Considering this
background curvature as an \textit{ambient} spacetime, the Christoffel
symbols and subsequently the Riemann tensor, become nonzero \(\left(\Gamma
^{\mu }{}_{\alpha \beta }\left(g_{\mu \nu }\right) \neq 0\right)\). Similarly,
the resulting compatibility conditions for the Cauchy deformation tensor states
that in Euclidean space the Riemann tensor is assigned to be
zero\cite{elastic1}; stating that local curvature is zero. But for the case of
non-euclidean geometries, this compatibility condition is extended to
pseudo-Riemannian geometries $\tilde{\Gamma }^{\mu }{}_{\alpha \beta
}\left(E_{\mu \nu }\right)\neq 0$. This contributes to the metric equivalence of
the Cauchy deformation tensor.

Using the changes in the spacetime metrics (Cauchy and background), large local
displacements described by an analogous Lagrange finite strain tensor in
$n=4$ dimensions can be constructed.
Here the finite spacetime strain, $\varepsilon_{\alpha\beta}$, is used as a
measure of how much a given deformation in the local spacetime differs from the
original configuration described by the background metric. For the most
fundamental case involving the minkowski metric for flat space the strain
tensor is represented as
\begin{equation}
\varepsilon _{\alpha \beta } = \frac{1}{2}\left(E_{\alpha \beta }-\eta _{\alpha
\beta }\right)
\end{equation}
For the case applicable to this work we use the strong-form background, where
the resulting strain is 
\begin{equation}
\varepsilon _{\alpha \beta } = \frac{1}{2}\left(E_{\alpha \beta
}-\Lambda_{\alpha \beta }\right)
\end{equation}
with the deformation line element as
\begin{equation}
\left(ds_1\right){}^2-\left(ds_0\right){}^2=2\varepsilon _{\mu \nu }(\eta ,\tau )d\eta ^{\nu }d\eta ^{\mu }
\end{equation}
where the parameters $\eta$ are the local coordinates of the
space and $\tau$ the proper time associated with the local spacetime. For an
arbitrarily rotating frame, the continuum line element of the space is
\begin{eqnarray} 
\left(ds_1\right){}^2-\left(ds_0\right){}^2&=&c^2\varepsilon _{00}dt^2-\varepsilon _{\text{ii}}\left(d\eta ^i\right){}^2\\\nonumber
&-&4\varepsilon_{\text{ij}}d\eta ^id\eta ^j\\\nonumber
&+&2\varepsilon_{\text{i0}}\text{cos$\gamma $}dtd\eta^i\nonumber
\end{eqnarray}
This strain tensor completely defines deformations of the resulting
{``}elastic{''} spacetime that it describes. In the following sections it will
be shown how this fundamental strain is used to define the overall gravitational
strain fields that describe deformations of mass-energy distributions contained
in a curved local spacetime, with nonzero scalar curvature and an introduced
constant of spactime elasticity; the Bulk Incompressibility ($\beta$) defined
in accordance to the $\Lambda$-CDM cosmological model.


\section{\label{fields}Gravitational Strain Tensor}

\subsection{\label{constants} Constant of Incompressibility and Resistance}
A bulk modulus as
a measure of the incompressibility of spacetime needs to be a resistive measure
of local contractions and bulk shearing in spacetime.
This modulus incorporates local and cosmological properties of the space, which
would include the use of a cosmological scale factor ($\alpha $($\tau $)) as a measure
of the local expansion of spacetime and densities associated with the
$\Lambda$-CDM background for the vacuum density, $\rho_{\text{vac}}$, and cold
dark matter, $\rho_{\text{CDM}}$. With these densities, we constrict an
approximation to consider these densities as homogeneous with respect to the
background measure. By including cosmological homogeneities that
arise, we can begin to construct a modulus constant that describes the resistive
nature of the cold dark matter cosmological background. With respect to this
background, this constant accompanies the nonlinear nature of the gravitational
field vacuum under a viscoelastic description as described in this work.
 
Beginning with the Friedmann equations for a monotonically expanding or
contracting universe\cite{mtw,carroll,hartle} we use cosmological
parameters to describe the resistive nature of the vacuum. This gives a much
more physically intuitive approach; as opposed to relying soley on a geometric
description in totality. Now, one could develop an expression for a bulk modulus
in terms of deriving the relative length contraction for a moving frame using
the Lorentz transformations; however, this method will ultimately cause
infinities to arise when applying the hyperbolic transformations to the energy
density as volume changes under rarefaction (compression) and dilation (contraction)\cite{rgdiss}. This
would give an indirect relationship of the relative change in volume due to a governing
pressure in the comoving frame. Along with this, a purely geometric
representation of the incompressibility will give an inappropriate description
of cosmological implications across cosmological distances. With that being
said, an expression of the incompressibility, in dimensions of pressure
$\left(M\cdot L^{-1}\cdot T^2\right)$, can be made via the Friedmann
equations and a \textit{volumetric velocity}, $\overset{\pmb{\cdot
}}{V}=V\left(3\frac{\overset{\pmb{\cdot }}{\alpha }}{\alpha }\right)$.
The square of the Hubble constant $H$ is approximated, neglecting the
contributions from non-gravitational radiation, extrinsic spatial curvature and
baryonic matter densities.
\begin{equation}
H^2=\frac{8\text{$\pi $G}}{3}\left(\rho _{\text{CDM}}+\rho _{\Lambda }\right)
\end{equation}
From this approximation, we can solve for the density of cold dark matter
$\left(\rho _{\text{CDM}}\right)$ in equation \ref{cdm} and make a substitution
for the vacuum density with respect to the time-dependent volume, $V(\tau)$,
and an effective vacuum mass, $m_{\text{vac}}$.
Where, $\tau$, is the cosmological time for the current epoch.
\begin{subequations}
\begin{eqnarray}\label{cdm}
\rho _{\text{CDM}}&=&\frac{1}{8\text{$\pi $G}}\left(3H^2-\text{$({}^{+}\Lambda)
$c}^2\right)\\
\rho _{\text{CDM}}&\equiv&\frac{dM_{\text{vac}}}{dV(\tau
)}=\frac{dM_{\text{vac}}}{\alpha ^3\sqrt{-g}dV}
\end{eqnarray}
\end{subequations}
Collecting all constants after rearrangement, including the dimensionless
scale factor ($\alpha$), we have
\begin{equation}
\frac{c^2}{\alpha ^3}\rho _{\text{CDM}}=-\frac{\sqrt{-g}}{8\text{$\pi
$G}}\left(3H^2-\text{$({}^{+}\Lambda) $c}^2\right)
\end{equation}
Lastly giving a \textit{Cosmological Bulk Incompressibility} constant ($\beta$)
of:
\begin{equation}\label{bulk}
\beta =-\frac{\sqrt{-g}}{\alpha ^38\text{$\pi
$G}}\left(3H^2c^2-\text{$({}^{+}\Lambda) $c}^4\right)
\end{equation}
This constant is suited for the description of a proposed value of the
inability of the dense $\Lambda $-CDM vacuum to be compressed.
Note that this resistance should always be negative corresponding to the
pressure felt due to the relative expansion of space. Combining this constant
with that of an effective vacuum mass ($m_{vac}$) that is a measure of the inertial
resistivity gives a formulation of a characteristic permitance of a
gravitational field; $\beta/2m_{vac}$. This formulation is loosely analogous to
that of the electromagnetic vacuum permittivity constant, $\epsilon_{0}$,
accompanying Maxwell's equations for the coupled description of electrodynamics.
With this, the modulus for bulk gravity is given by equation \ref{bulk}
using values of cosmological
parameters in H and ${}^{+}\Lambda$\cite{cosmo1,wmap1}, where an approximated
value of ($\beta$) is:
\begin{eqnarray}
\beta &=&-\frac{\sqrt{-det[\Lambda_{\mu\nu}]}}{\alpha^{3}8\text{$\pi
$G}}\left(3H^{2}c^{2}-\text{$({}^{+}\Lambda) $}c^{4}\right)\\\nonumber
&=& 8.30977\times10^{29} \text{$N\cdot m^{-2}$}
\end{eqnarray}


\subsection{Field Tensor Components}

From this definition of \textit{spacetime strain} and a modulus for describing
the resistive nature of the $\Lambda$-CDM vacuum we now introduce the
resulting \textit{Gravitational Strain Fields}. Analyzing the fundamental
components of the strain tensor explained in previous sections of this work, we
can parse these terms and construct corresponding gravitational strain
fields; fundamentally based upon the elastic nature of spacetime. Thus, creating
two different tensor fields describing volume (longitudinal) and shear
(tangential) deformations.
These, respectfully, are the \textit{Dilation} $(\mathcal{D}^{\mu\nu})$
\begin{equation}
\mathcal{D}^{\mu\nu} =\frac{\beta
R_{vac}}{2m_{vac}}\Lambda^{\mu\nu}\left(\Lambda_{\sigma\gamma}\varepsilon^{\sigma\gamma}\right)
\end{equation}
and \textit{Shear-Tidal} $(\mathcal{S}^{\mu\nu})$ tensor fields
\begin{equation}
\mathcal{S}^{\mu\nu}=\frac{\nu_{irr}R_{vac}}{2m_{vac}}\left[\varepsilon^{\mu
\nu }-\frac{1}{4}\Lambda^{\mu \nu }\left(\Lambda _{\sigma
\gamma}\varepsilon^{\sigma \gamma}\right)\right]
\end{equation}

These separated fields can then be recoupled or repackaged together to form the
gravitational strain field tensor, analogous to the Faraday tensor for the
electromagnetic field. The table below gives a statement of the symmetries of
the coupled gravitational strain fields, represented in terms of $\Psi^{\mu\nu}$, where ($c_g$)
is the propagation speed of the fields in gravitational vacuum. Much like the
properties of the electromagnetic field tensor (or Faraday tensor) $F^{\mu\nu}$,
with the exception of the trace being nonzero for $\Psi^{\mu\nu}$. The inclusion
of diagonal terms for this strain tensor
\begingroup
\begin{table}[h]
\centering
\begin{align*}
\boxed{
\Psi^{\mu\nu}
\begin{dcases}
 \text{Symmetry}:  & \Psi ^{(\mu \nu)} =\frac{1}{2}\left(\Psi^{\nu \mu
 }+\Psi^{\nu\mu}\right)\\
  \text{Antisymmetry}:
 & \Psi^{[\mu\nu]} =\frac{1}{2}\left(\Psi^{\mu\nu}-\Psi^{\nu\mu}\right)\\
\text{Inner Product}: & \Psi _{\mu \nu }\Psi ^{\mu \nu } =
 \left(\mathcal{D}_{00}\right)^2+3\left(\mathcal{D}_{ii}\right)^2\\
 &\qquad \qquad{ }  -\frac{2}{c^2}\mathcal{S}_{\mu\nu}\mathcal{S}^{\mu\nu}
\end{dcases}
}
\end{align*}
\caption{Symmetries of the skew-symmetric Gravitational Strain
Field tensor, ($\Psi^{\mu\nu}$).}
\end{table}
\endgroup
 
For $\Psi^{\mu\nu}$ in terms of only the shear-tidal and
dilation fields, the diagonal elements represent the longitudinal components of
the general tensor field in terms of the dilation field, $\mathcal{D}^{\mu\nu}$.
The off-diagonal elements of $\Psi^{\mu\nu}$ are the
components of the shear strain of the field, where (n) is the dimensionality of
the space and ($\nu_{\text{irr}} $) is the irrotational \textit{viscosity}
constant associated with viscous deformations. Without rotations about the
spatial axes, the shear-tidal field is expressed as
\begin{eqnarray}
\Psi^{(\mu\nu)
}&\equiv&\mathcal{S}^{\mu\nu}\\
&=&\frac{2n+2}{n(n+1)}\frac{\nu_{irr}R_{vac}}{m_{vac}}\left[\varepsilon^{\mu
\nu }-\frac{1}{n}\Lambda^{\mu \nu }\left(\Lambda _{\sigma
\gamma}\varepsilon^{\sigma \gamma}\right)\right]\nonumber
\end{eqnarray}
Where the components $\Psi^{\text{i0}}$ of the tensor are the
antiplanar strain very much like the linearized Einstein
field equations in the Transverse-Traceless-gauge \cite{ttgauge}. The dilation
field is expressed as the trace of the gravitational strain tensor.
\begin{eqnarray}
\Lambda^{\mu\nu}\left(\Lambda_{\sigma \gamma
 }\Psi ^{\sigma \gamma }\right)&\equiv&\mathcal{D}^{\mu\nu}\\ 
 &=&\frac{2}{n}
 \frac{\beta
 R_{vac}}{m_{vac}}\Lambda^{\mu\nu}\left(\Lambda_{\sigma\gamma}\varepsilon^{\sigma\gamma}\right)\nonumber
\end{eqnarray}
When rotations are considered, we add to the spatial
components of $\mathcal{S}$, a \textit{viscous} shear due to rotations of local
spacetime; generating the antisymmetric components $\mathcal{S}^{[ij]}$.
\begin{eqnarray}
\Psi ^{\mu \nu }&=& \left(
\begin{array}{cccc}
 -\mathcal{D}^{00} & -\mathcal{S}^{10} & -\mathcal{S}^{20} & -\mathcal{S}^{30} \\
 \mathcal{S}^{01} & \mathcal{D}^{11} & -\mathcal{S}^{21} & \mathcal{S}^{31} \\
 \mathcal{S}^{02} & \mathcal{S}^{12} & \mathcal{D}^{22} & -\mathcal{S}^{32} \\
 \mathcal{S}^{03} & -\mathcal{S}^{13} & \mathcal{S}^{23} & \mathcal{D}^{33} \\
\end{array}
\right)\\
&=&\left(
\begin{array}{cccc}
 0 & -\mathcal{S} ^{10} & -\mathcal{S} ^{20} & -\mathcal{S} ^{30} \\
 \mathcal{S} ^{01} & 0 & -\mathcal{S} ^{21} & \mathcal{S} ^{31} \\
 \mathcal{S} ^{02} & \mathcal{S} ^{12} & 0 & -\mathcal{S} ^{32} \\
 \mathcal{S} ^{03} & -\mathcal{S} ^{13} & \mathcal{S} ^{23} & 0 \\
\end{array}
\right)\\\nonumber &+& \left(
\begin{array}{cccc}
 -\mathcal{D}^{00} & 0 & 0 & 0 \\
 0 & \mathcal{D}^{11} & 0 & 0 \\
 0 & 0 & \mathcal{D}^{22} & 0 \\
 0 & 0 & 0 & \mathcal{D}^{33} \\
\end{array}\right)\nonumber
\end{eqnarray}
As can be seen, the total strain is comprised of two coupled fields. Much like
the electric field, the gravitational field behaves in the same
manner. Thus, it is then logical to associate a symmetric (non-rotating) tensor
representation of the field\cite{rdecomp1,maxwell1}. As for a complete
solenoidal (rotating) field for gravitation, an elusive magnetic analog concerning a
proposed gravitational \textit{vorticity} is not so intuitive. The table below
groups these coupled fields in terms of the respective transverse and
longitudinal parts of the gravitational strain tensor.
\begin{table}[h]
\centering
\fbox{$\Psi^{\mu \nu }
\begin{cases}
 \text{Irrotational Field}: & \mathcal{S}^{0i}\equiv c_g\Psi^{0i}\\
 \text{Solenoidal Part}: & \mathcal{S}^{ij}\equiv \Psi^{[ij]},\text{ } \text{for
 i}\neq\text{j}\\
 \text{Longitudinal Field}: &
 \mathcal{D}^{\mu\nu}\equiv\Lambda^{\mu\nu}\left(\Lambda _{\sigma \gamma }\Psi ^{\sigma \gamma }\right)
\end{cases}
$}
\caption{Components and properties of the skew-symmetric Gravitational Strain
Field tensor, ($\Psi^{\mu\nu}$).}
\end{table}



\section{Interpreting \textit{Gravitational} Strain}

Realizing volume deformations of source mass distributions, the dilation field
constitutes deformations when considering the principle directions or
local volumes. In terms of gravitational physics, mass densities and scalar
curvatures are considered the sources of this longitudinal field. Under static
conditions this scalar-valued tensor field is described by the finite volumetric gravitational strain for a
constant local mass distribution. Expansion of the dilation
tensor is given in terms of the finite volumetric strain and the trace of the
background vacuum Ricci curvature tensor, the Ricci scalar ($R_{vac}$), for the
local spacetime.
\begin{equation}
\mathcal{D}^{\mu\nu} =\frac{\beta
R_{vac}}{2m_{vac}}\Lambda^{\mu\nu}\left(\Lambda_{\sigma\gamma}\varepsilon^{\sigma\gamma}\right)
\end{equation}
In this expression the vacuum scalar curvature is provided by the Einstein field
equations of the strong-form background, in which the scalar holds
values for the background vacuum ($\Lambda_{\mu\nu}$) and the source
distribution.
When considering the ${}^{+}\Lambda$-vacuum as a zero state configuration, the scalar curvature is always nonzero with a value of
$R_{vac}\equiv R_{zero}$, when $R_{zero} =
\Lambda^{\alpha\sigma}R_{\alpha\sigma}\equiv 4({}^{+}\Lambda)$ for dimension (n
= 4). For nonzero source distribution, $R_{vac} = R_{zero} + R_{source}$ where
$R_{source}$ can take on values from any solution of the trace Einstein field
equations for nonzero mass.
This equation for the dilation field represents a way of dynamically describing
the gravitational strain field generated by a nonzero mass-energy distribution
using the trace Einstein field equations; as previously explained.

The Shear-Tidal field represents irrotational shape or surface deformations of a
mass distribution. In relation to gravitational field theory,
the shear-tidal part represents the tidal field generated by a source
of gravitational stress acting upon a central mass. This is expressed as the
following for reiteration:
\begin{equation}
\mathcal{S}^{\mu\nu}=\frac{\beta R_{vac}}{2m_{vac}}\left[\varepsilon^{\mu
\nu }-\frac{1}{4}\Lambda^{\mu \nu }\left(\Lambda _{\sigma
\gamma}\varepsilon^{\sigma \gamma}\right)\right]
\end{equation}
Where, again for reiteration, the components
$\left(\mathcal{S}^{\text{i0}}\right)$ of the tensor are the anti-planar strain much like the linearized Einstein field equations. The
scalar curvature and effective mass of the background vacuum are represented as
previously explained, with the exception of the field
constant in $(\nu_{\text{irr}})$. This constant is a representation of the
viscous-like flow that accompanies the shearing deformations of distributions.
On a cosmological scale, the irrotational permittivity of a gravitational field
is governed by the bulk field constant, ($\beta$). Over these cosmological
distances, the bulk field constant is the only significant parameter that
regulates this proposed permittivity of a gravitational field.

\section{Conclusions}

Having established a dynamic background metric tensor in the first section
allows for a non-trivial use of the metric tensor as a reference measure.
Using the background metric (equations \ref{eq:strong} and
\ref{eq:weak}) in this fashion aides in the design of the overall field
theory.

Historically, as is explained above, using a
purely geometrical (mathematical) theory of gravity excludes a notion or
intuition of implementing empirical experiments that test for the dynamics
outlined in a metric theory of gravity \cite{observe1,observe3}.
Currently in the field of gravitational theory, there is no consistent theory of
gravitation that produces time-dependent solutions that also admit a wave nature
for gravitational interactions. This work presented seeks to extend our
interpretations of gravity as a fundamental force, such that it can then be
used to aid a more robust formulation of the wave nature of the field.
With the axioms of the vacuum fluidity in section \ref{axiom}, the
background spacetime naturally gives rise to properties of elasticity in
4-dimensions. The next steps are to formulate the full rotational contributions
to this viscoelastic representation of gravitation. While exploring variations
of these fields; when investigating a formal Lagrangian density and respective
action integral.


\bibliographystyle{apsrev4-1}

\begin{thebibliography}{20}%
\makeatletter
\providecommand \@ifxundefined [1]{%
 \@ifx{#1\undefined}
}%
\providecommand \@ifnum [1]{%
 \ifnum #1\expandafter \@firstoftwo
 \else \expandafter \@secondoftwo
 \fi
}%
\providecommand \@ifx [1]{%
 \ifx #1\expandafter \@firstoftwo
 \else \expandafter \@secondoftwo
 \fi
}%
\providecommand \natexlab [1]{#1}%
\providecommand \enquote  [1]{``#1''}%
\providecommand \bibnamefont  [1]{#1}%
\providecommand \bibfnamefont [1]{#1}%
\providecommand \citenamefont [1]{#1}%
\providecommand \href@noop [0]{\@secondoftwo}%
\providecommand \href [0]{\begingroup \@sanitize@url \@href}%
\providecommand \@href[1]{\@@startlink{#1}\@@href}%
\providecommand \@@href[1]{\endgroup#1\@@endlink}%
\providecommand \@sanitize@url [0]{\catcode `\\12\catcode `\$12\catcode
  `\&12\catcode `\#12\catcode `\^12\catcode `\_12\catcode `\%12\relax}%
\providecommand \@@startlink[1]{}%
\providecommand \@@endlink[0]{}%
\providecommand \url  [0]{\begingroup\@sanitize@url \@url }%
\providecommand \@url [1]{\endgroup\@href {#1}{\urlprefix }}%
\providecommand \urlprefix  [0]{URL }%
\providecommand \Eprint [0]{\href }%
\providecommand \doibase [0]{http://dx.doi.org/}%
\providecommand \selectlanguage [0]{\@gobble}%
\providecommand \bibinfo  [0]{\@secondoftwo}%
\providecommand \bibfield  [0]{\@secondoftwo}%
\providecommand \translation [1]{[#1]}%
\providecommand \BibitemOpen [0]{}%
\providecommand \bibitemStop [0]{}%
\providecommand \bibitemNoStop [0]{.\EOS\space}%
\providecommand \EOS [0]{\spacefactor3000\relax}%
\providecommand \BibitemShut  [1]{\csname bibitem#1\endcsname}%
\let\auto@bib@innerbib\@empty
\bibitem [{\citenamefont {Abbott}(2016)}]{ligo}%
  \BibitemOpen
  \bibfield  {author} {\bibinfo {author} {\bibfnamefont {B.~P. e.~a.}\
  \bibnamefont {Abbott}},\ }\href {\doibase 10.1103/PhysRevLett.116.061102}
  {\bibfield  {journal} {\bibinfo  {journal} {Phys. Rev. Lett.}\ }\textbf
  {\bibinfo {volume} {116}},\ \bibinfo {pages} {061102} (\bibinfo {year}
  {2016})},\ \bibinfo {note} {lIGO Scientific Collaboration and Virgo
  Collaboration}\BibitemShut {NoStop}%
\bibitem [{\citenamefont {Martynov}\ \emph {et~al.}(2016)\citenamefont
  {Martynov} \emph {et~al.}}]{ligo2}%
  \BibitemOpen
  \bibfield  {author} {\bibinfo {author} {\bibfnamefont {D.~V.}\ \bibnamefont
  {Martynov}} \emph {et~al.} (\bibinfo {collaboration} {LIGO Scientific}),\
  }\href {\doibase 10.1103/PhysRevD.93.112004} {\bibfield  {journal} {\bibinfo
  {journal} {Phys. Rev.}\ }\textbf {\bibinfo {volume} {D93}},\ \bibinfo {pages}
  {112004} (\bibinfo {year} {2016})},\ \Eprint
  {http://arxiv.org/abs/1604.00439} {arXiv:1604.00439 [astro-ph.IM]}
  \BibitemShut {NoStop}%
\bibitem [{\citenamefont {ESA/ATG~medialab}(2015)}]{lisa}%
  \BibitemOpen
  \bibfield  {author} {\bibinfo {author} {\bibfnamefont {E.}~\bibnamefont
  {ESA/ATG~medialab}},\ }\href
  {http://www.esa.int/Our_Activities/Space_Science/LISA} {\enquote {\bibinfo
  {title} {Lisa pathfinder mission},}\ } (\bibinfo {year} {2015})\BibitemShut
  {NoStop}%
\bibitem [{\citenamefont {{Weisberg}}\ \emph {et~al.}(1981)\citenamefont
  {{Weisberg}}, \citenamefont {{Taylor}},\ and\ \citenamefont
  {{Fowler}}}]{pulsar}%
  \BibitemOpen
  \bibfield  {author} {\bibinfo {author} {\bibfnamefont {J.~M.}\ \bibnamefont
  {{Weisberg}}}, \bibinfo {author} {\bibfnamefont {J.~H.}\ \bibnamefont
  {{Taylor}}}, \ and\ \bibinfo {author} {\bibfnamefont {L.~A.}\ \bibnamefont
  {{Fowler}}},\ }\href {\doibase 10.1038/scientificamerican1081-74} {\bibfield
  {journal} {\bibinfo  {journal} {Scientific American}\ }\textbf {\bibinfo
  {volume} {245}},\ \bibinfo {pages} {74} (\bibinfo {year} {1981})}\BibitemShut
  {NoStop}%
\bibitem [{\citenamefont {Einstein}(1915)}]{efe}%
  \BibitemOpen
  \bibfield  {author} {\bibinfo {author} {\bibfnamefont {A.}~\bibnamefont
  {Einstein}},\ }\href@noop {} {\bibfield  {journal} {\bibinfo  {journal}
  {Preussische Akademie der Wissenschaften, Sitzungsberichte}\ } (\bibinfo
  {year} {1915})}\BibitemShut {NoStop}%
\bibitem [{\citenamefont {Aldrovandi}\ \emph {et~al.}(2007)\citenamefont
  {Aldrovandi}, \citenamefont {Pereira},\ and\ \citenamefont {Vu}}]{nonlinear}%
  \BibitemOpen
  \bibfield  {author} {\bibinfo {author} {\bibfnamefont {R.}~\bibnamefont
  {Aldrovandi}}, \bibinfo {author} {\bibfnamefont {J.~G.}\ \bibnamefont
  {Pereira}}, \ and\ \bibinfo {author} {\bibfnamefont {K.~H.}\ \bibnamefont
  {Vu}},\ }\href {\doibase 10.1007/s10701-007-9180-2} {\bibfield  {journal}
  {\bibinfo  {journal} {Foundations of Physics}\ }\textbf {\bibinfo {volume}
  {37}},\ \bibinfo {pages} {1503} (\bibinfo {year} {2007})}\BibitemShut
  {NoStop}%
\bibitem [{\citenamefont {Yunes}\ and\ \citenamefont
  {Siemens}(2013)}]{pulsar2}%
  \BibitemOpen
  \bibfield  {author} {\bibinfo {author} {\bibfnamefont {N.}~\bibnamefont
  {Yunes}}\ and\ \bibinfo {author} {\bibfnamefont {X.}~\bibnamefont
  {Siemens}},\ }\href {\doibase 10.12942/lrr-2013-9} {\bibfield  {journal}
  {\bibinfo  {journal} {Living Rev. Rel.}\ }\textbf {\bibinfo {volume} {16}},\
  \bibinfo {pages} {9} (\bibinfo {year} {2013})},\ \Eprint
  {http://arxiv.org/abs/1304.3473} {arXiv:1304.3473 [gr-qc]} \BibitemShut
  {NoStop}%
\bibitem [{\citenamefont {Kip S.~Thorne}(1973)}]{mtw}%
  \BibitemOpen
  \bibfield  {author} {\bibinfo {author} {\bibfnamefont {C.~W.~M.}\
  \bibnamefont {Kip S.~Thorne}, \bibfnamefont {John A.~Wheeler}},\ }\href@noop
  {} {\emph {\bibinfo {title} {Gravitation}}}\ (\bibinfo  {publisher} {W. H.
  Freeman},\ \bibinfo {year} {1973})\BibitemShut {NoStop}%
\bibitem [{\citenamefont {Muller}\ and\ \citenamefont
  {Grave}(2009)}]{spacetime}%
  \BibitemOpen
  \bibfield  {author} {\bibinfo {author} {\bibfnamefont {T.}~\bibnamefont
  {Muller}}\ and\ \bibinfo {author} {\bibfnamefont {F.}~\bibnamefont {Grave}},\
  }\href@noop {} {\  (\bibinfo {year} {2009})},\ \Eprint
  {http://arxiv.org/abs/0904.4184} {arXiv:0904.4184 [gr-qc]} \BibitemShut
  {NoStop}%
\bibitem [{\citenamefont {Marsden}\ and\ \citenamefont
  {Hughes}(1994)}]{elastic1}%
  \BibitemOpen
  \bibfield  {author} {\bibinfo {author} {\bibfnamefont {J.}~\bibnamefont
  {Marsden}}\ and\ \bibinfo {author} {\bibfnamefont {T.}~\bibnamefont
  {Hughes}},\ }\href {https://books.google.com/books?id=RjzhDL5rLSoC} {\emph
  {\bibinfo {title} {Mathematical Foundations of Elasticity}}},\ Dover Civil
  and Mechanical Engineering Series\ (\bibinfo  {publisher} {Dover},\ \bibinfo
  {year} {1994})\BibitemShut {NoStop}%
\bibitem [{\citenamefont {Carroll}(2003)}]{carroll}%
  \BibitemOpen
  \bibfield  {author} {\bibinfo {author} {\bibfnamefont {S.~M.}\ \bibnamefont
  {Carroll}},\ }\href@noop {} {\emph {\bibinfo {title} {Spacetime and
  Geometry}}}\ (\bibinfo  {publisher} {Pearson/Addison Wesley},\ \bibinfo
  {year} {2003})\BibitemShut {NoStop}%
\bibitem [{\citenamefont {Hartle}(2003)}]{hartle}%
  \BibitemOpen
  \bibfield  {author} {\bibinfo {author} {\bibfnamefont {J.~B.}\ \bibnamefont
  {Hartle}},\ }\href@noop {} {\emph {\bibinfo {title} {Gravity}}}\ (\bibinfo
  {publisher} {Pearson/Addison Wesley},\ \bibinfo {year} {2003})\BibitemShut
  {NoStop}%
\bibitem [{\citenamefont {Gamble~Jr}(2017)}]{rgdiss}%
  \BibitemOpen
  \bibfield  {author} {\bibinfo {author} {\bibfnamefont {R.~S.}\ \bibnamefont
  {Gamble~Jr}},\ }\href@noop {} {\emph {\bibinfo {title} {On Gravitational
  Radiation: A Nonlinear Wave Theory In A Viscoelastic Kerr-$\Lambda$
  Spacetime}}}\ (\bibinfo  {publisher} {ProQuest, North Carolina Agricultural
  \& Technical State University},\ \bibinfo {year} {2017})\BibitemShut
  {NoStop}%
\bibitem [{\citenamefont {Bernabeu}\ \emph {et~al.}(2011)\citenamefont
  {Bernabeu}, \citenamefont {Espriu},\ and\ \citenamefont
  {Puigdom\`enech}}]{cosmo1}%
  \BibitemOpen
  \bibfield  {author} {\bibinfo {author} {\bibfnamefont {J.}~\bibnamefont
  {Bernabeu}}, \bibinfo {author} {\bibfnamefont {D.}~\bibnamefont {Espriu}}, \
  and\ \bibinfo {author} {\bibfnamefont {D.}~\bibnamefont {Puigdom\`enech}},\
  }\href {\doibase 10.1103/PhysRevD.84.063523} {\bibfield  {journal} {\bibinfo
  {journal} {Phys. Rev. D}\ }\textbf {\bibinfo {volume} {84}},\ \bibinfo
  {pages} {063523} (\bibinfo {year} {2011})}\BibitemShut {NoStop}%
\bibitem [{\citenamefont {{Jarosik}}\ and\ \citenamefont
  {{Bennett}}(2011)}]{wmap1}%
  \BibitemOpen
  \bibfield  {author} {\bibinfo {author} {\bibfnamefont {N.}~\bibnamefont
  {{Jarosik}}}\ and\ \bibinfo {author} {\bibfnamefont {C.~L.~e.}\ \bibnamefont
  {{Bennett}}},\ }\href {\doibase 10.1088/0067-0049/192/2/14}{\bibfield {journal} {\bibinfo  {journal} {apjs}\ }\textbf {\bibinfo {volume}
  {192}  (\bibinfo {year} {2011})}},\ \Eprint
  {http://arxiv.org/abs/1001.4744} {arXiv:1001.4744}\BibitemShut {NoStop}%
\bibitem [{\citenamefont {Teukolsky}(1982)}]{ttgauge}%
  \BibitemOpen
  \bibfield  {author} {\bibinfo {author} {\bibfnamefont {S.~A.}\ \bibnamefont
  {Teukolsky}},\ }\href {\doibase 10.1103/PhysRevD.26.745} {\bibfield
  {journal} {\bibinfo  {journal} {Phys. Rev. D}\ }\textbf {\bibinfo {volume}
  {26}},\ \bibinfo {pages} {745} (\bibinfo {year} {1982})}\BibitemShut
  {NoStop}%
\bibitem [{\citenamefont {Schmidt}(2004)}]{rdecomp1}%
  \BibitemOpen
  \bibfield  {author} {\bibinfo {author} {\bibfnamefont {H.-J.}\ \bibnamefont
  {Schmidt}},\ }\href@noop {} {\  (\bibinfo {year} {2004})},\ \Eprint
  {http://arxiv.org/abs/gr-qc/0407053} {arXiv:gr-qc/0407053 [gr-qc]}
  \BibitemShut {NoStop}%
\bibitem [{\citenamefont {{Hauser}}(1970)}]{maxwell1}%
  \BibitemOpen
  \bibfield  {author} {\bibinfo {author} {\bibfnamefont {W.}~\bibnamefont
  {{Hauser}}},\ }\href {\doibase 10.1119/1.1976233} {\bibfield  {journal}
  {\bibinfo  {journal} {American Journal of Physics}\ }\textbf {\bibinfo
  {volume} {38}},\ \bibinfo {pages} {80} (\bibinfo {year} {1970})}\BibitemShut
  {NoStop}%
\bibitem [{\citenamefont {Rovelli}(2002)}]{observe1}%
  \BibitemOpen
  \bibfield  {author} {\bibinfo {author} {\bibfnamefont {C.}~\bibnamefont
  {Rovelli}},\ }\href {\doibase 10.1103/PhysRevD.65.124013} {\bibfield
  {journal} {\bibinfo  {journal} {Phys. Rev.}\ }\textbf {\bibinfo {volume}
  {D65}},\ \bibinfo {pages} {124013} (\bibinfo {year} {2002})},\ \Eprint
  {http://arxiv.org/abs/gr-qc/0110035} {arXiv:gr-qc/0110035 [gr-qc]}
  \BibitemShut {NoStop}%
\bibitem [{\citenamefont {Lusanna}\ and\ \citenamefont
  {Pauri}(2006)}]{observe3}%
  \BibitemOpen
  \bibfield  {author} {\bibinfo {author} {\bibfnamefont {L.}~\bibnamefont
  {Lusanna}}\ and\ \bibinfo {author} {\bibfnamefont {M.}~\bibnamefont
  {Pauri}},\ }\href {\doibase 10.1007/s10714-005-0218-5} {\bibfield  {journal}
  {\bibinfo  {journal} {Gen. Rel. Grav.}\ }\textbf {\bibinfo {volume} {38}},\
  \bibinfo {pages} {229} (\bibinfo {year} {2006})},\ \Eprint
  {http://arxiv.org/abs/gr-qc/0407007} {arXiv:gr-qc/0407007 [gr-qc]}
  \BibitemShut {NoStop}%
\end{thebibliography}

%


\end{document}